\documentclass[twocolumn,showpacs,preprintnumbers,amsmath,amssymb]{revtex4}

\usepackage{graphicx}
\usepackage{dcolumn}
\usepackage{bm}
\usepackage{dsfont}
\usepackage{mathrsfs}
\usepackage{amsmath}				

\newcommand{\be}{\begin{equation}}
\newcommand{\ee}{\end{equation}}
\newcommand{\bey}{\begin{eqnarray}}
\newcommand{\eey}{\end{eqnarray}}

\begin{document} 
\title{Dual lattice simulation of the U(1) gauge-Higgs model at finite density  \\
- an exploratory  proof-of-concept study}
\author{Ydalia Delgado Mercado, Christof Gattringer, Alexander Schmidt}
\affiliation{
\vspace{3mm}
Institut f\"ur Physik, FB Theoretische Physik,
Universit\"at Graz, 8010 Graz, Austria}

\begin{abstract}
The U(1) gauge-Higgs model with two flavors of opposite charge and a chemical potential is mapped exactly to a
dual representation where  matter fields correspond to loops of flux and the gauge fields are represented
by surfaces. The complex action problem of the  conventional formulation at finite  chemical potential
$\mu$ is overcome in the dual representation and the partition sum has only real and non-zero contributions. We
simulate the  model in the dual representation using a generalized worm algorithm, explore the phase
diagram and study condensation phenomena at finite $\mu$. 
\end{abstract}

\maketitle

\noindent
{\bf Introductory remarks: }
In recent years lattice QCD has turned into a powerful quantitative tool in hadron physics.
However, one aspect where lattice methods still face serious technical obstacles is QCD at
finite density. The reason is that at finite chemical potential $\mu$ the action $S$ is
complex  and the Boltzmann factor $e^{-S}$ cannot be used as weight factor in a Monte Carlo
simulation. 

For some lattice models considerable progress was made by mapping the system to new (dual)
degrees of freedom, where the partition  sum has only real and positive terms (see
\cite{examples0,examples1,examples2,examples3,examples4,examples5,DeGaSch1,phi4spectro} for examples related to this
work). The dual variables are typically fluxes on the lattice that are subject to constraints. The
worm algorithm \cite{worm} is a  powerful tool for updating such constrained systems.

In this Letter we present a first proof-of-concept study for a system with gauge and matter fields at arbitrary
couplings and finite density. We consider the U(1) gauge-Higgs model
with two flavors and chemical potential. The corresponding
dual representation is given in terms of closed loops of flux for the matter fields and
surfaces for the gauge fields. For the Monte Carlo we compare two techniques and 
show that the dual approach successfully overcomes the complex action problem. We explore the
phase diagram and as illustrative examples discuss Silver-Blaze type of transitions 
\cite{cohen} and show that they
can be understood as condensation of the dual variables.

\vskip2mm

\noindent  
{\bf U(1) gauge-Higgs model on the lattice:}
We here consider the model with two flavors of opposite charge described
by complex scalar fields $\phi_x, \chi_x \in \mathds{C}$ living on the 
sites $x$ of the lattice. The gauge fields $U_{x,\sigma} \in$ U(1) live on the links. Throughout 
this paper we use 4-d euclidean lattices of size $V_4 = N_s^3 \times N_t$ with periodic 
boundary conditions for all directions. The lattice spacing is set to 1, i.e., all dimensionful quantities 
are in units of the lattice spacing. Scale setting can be implemented as in any other lattice field theory 
and issues concerning the continuum behavior are, e.g., discussed in \cite{LuWe}.
We write the action as the sum, 
$S = S_U + S_\phi + S_\chi$, where $S_U$ is the gauge action and $S_\phi$ and $S_\chi$ are the actions for the two scalars. 
For the gauge action we use 
Wilson's form
\begin{equation} 
S_U \; = \; - \beta \, \sum_x \sum_{\sigma < \tau} \mbox{Re} \; U_{x,\sigma} U_{x+\widehat{\sigma}, \tau}
U_{x+\widehat{\tau},\sigma}^\star U_{x,\tau}^\star \; .
\label{gaugeaction}
\end{equation}
The sum runs over all plaquettes, $\widehat{\sigma}$ and $\widehat{\tau}$ denote the unit vectors in $\sigma$- and 
$\tau$-direction and the asterisk is used for complex conjugation.  
The action for the field $\phi$ is 
\begin{eqnarray}
&& \qquad S_\phi   
=  \! \sum_x \!\Big(  M_\phi^2 \, |\phi_x|^2  + \lambda_\phi |\phi_x|^4  -
\label{matteraction} \\
&& \sum_{\nu = 1}^4 \!
\big[  e^{-\mu_\phi  \delta_{\nu, 4} } \, \phi_x^\star \, U_{x,\nu} \,\phi_{x+\widehat{\nu}} 
\, + \, 
e^{\mu_\phi \delta_{\nu, 4}} \, \phi_x^\star \, 
U_{x-\widehat{\nu}, \nu}^\star \, \phi_{x-\widehat{\nu}}  \big] \!  \Big) .
\nonumber
\end{eqnarray}
By $M_\phi^2$  we denote the combination $8 + m_\phi^2$, where $m_\phi$ is the bare mass
parameter of the field $\phi$ and $\mu_\phi$ is the chemical potential, which favors forward
hopping in time-direction (= 4-direction). The coupling for the quartic term is denoted as 
$\lambda_\phi$. The action for the field $\chi$ has the same form as
(\ref{matteraction}) but with complex conjugate link variables $U_{x,\nu}$ such that $\chi$ has
opposite charge.  $M_\chi^2$, $\mu_\chi$ and $\lambda_\chi$  are used for the parameters of $\chi$. 

The partition sum $Z = \int D[U] D[\phi,\chi] e^{-S_U - S_\chi - S_\phi}$  is obtained by
integrating the Boltzmann factor over all field configurations. The measures are products over
the measures for each individual degree of freedom.

Note that for $\mu_\phi \neq 0$ (\ref{matteraction}) is complex, i.e., in the
conventional form  the theory has a complex action problem.

\vskip2mm
\noindent  
{\bf Dual representation:} A detailed derivation of the dual representation for the 1-flavor
model is given in \cite{DeGaSch1} and the generalization to two flavors is straightforward.
The final result 
for the dual representation of the partition sum for the gauge-Higgs model with two flavors is
\begin{equation}
\hspace*{-3mm} Z = \!\!\!\!\!\! \sum_{\{p,j,\overline{j},l,\overline{l} \}} \!\!\!\!\!\!  {\cal C}_g[p,j,l]  \;  {\cal C}_s  [j] \;   {\cal C}_s  [l] \;  {\cal W}_U[p] 
\; {\cal W}_\phi \big[j,\overline{j}\,\big] \, {\cal W}_\chi \big[l,\overline{l}\,\big]  .
\label{Zfinal}
\end{equation} 
The sum runs over all configurations of the dual variables: The occupation numbers 
$p_{x,\sigma\tau} \in \mathds{Z}$ assigned to the plaquettes of the lattice and the flux variables  $j_{x,\nu},  l_{x,\nu} \in \mathds{Z}$ and
$\overline{j}_{x,\nu},  \overline{l}_{x,\nu} \in \mathds{N}_0$ living on the links. The flux variables $j$ and $l$ are subject
to the constraints ${\cal C}_s$ (here $\delta(n)$ denotes the Kronecker delta $\delta_{n,0}$ and $\partial_\nu f_x \equiv 
f_x - f_{x-\widehat{\nu}}$)
\begin{equation}
 {\cal C}_s [j] \, = \, \prod_x \delta \! \left( \sum_\nu \partial_\nu j_{x,\nu}  \right) , \;
\label{loopconstU1}
\end{equation}
which enforce the conservation of $j$-flux and of $l$-flux at each site of the lattice.
Another constraint,
\begin{equation}
 {\cal C}_g [p,j,l]  \! =\!  \prod_{x,\nu} \! \delta  \Bigg( \!\sum_{\nu < \alpha}\! \partial_\nu p_{x,\nu\alpha}  
- \!\sum_{\alpha<\nu}\! \partial_\nu p_{x,\alpha\nu} + j_{x,\nu} - l_{x,\nu} \! \Bigg)\! ,
\label{plaqconstU1}  
\end{equation}
connects the plaquette occupation numbers $p$ with the $j$- and $l$-variables. 
At every link it enforces the combined flux of the plaquette occupation 
numbers  plus the difference of $j$- and $l$-flux residing on that link to vanish. 

The constraints (\ref{loopconstU1}) and (\ref{plaqconstU1}) restrict the admissible
flux and plaquette occupation numbers giving rise to an interesting geometrical
interpretation: The $j$- and $l$-fluxes form closed oriented loops made of links. The integers
$j_{x,\nu}$ and $l_{x,\nu}$ determine how often a link is run through by loop segments, with negative
numbers indicating net flux in the negative direction. The flux conservation 
(\ref{loopconstU1}) ensures that only closed loops appear. Similarly, the constraint 
(\ref{plaqconstU1}) for the plaquette occupation numbers can be seen as a continuity
condition for surfaces made of plaquettes. The surfaces are either closed
surfaces without boundaries or open surfaces bounded by  $j$- or $l$-flux.

The configurations of plaquette occupation numbers and fluxes in (\ref{Zfinal}) come with 
weight factors 
\begin{eqnarray}
{\cal W}_U[p] & = & \!\! \! \prod_{x,\sigma < \tau} \! \! \!
 I_{p_{x,\sigma\tau}}(\beta) \, ,
\\   
{\cal W}_\phi \big[j,\overline{j}\big] & = & 
\prod_{x,\nu}\! \frac{1}{(|j_{x,\nu}|\! +\! \overline{j}_{x,\nu})! \, 
\overline{j}_{x,\nu}!} 
\prod_x e^{-\mu j_{x,4}}  P_\phi \left( f_x \right) ,
\nonumber
\end{eqnarray}
with $f_x = \sum_\nu\!\big[ |j_{x,\nu}|\!+\!  |j_{x-\widehat{\nu},\nu}|  \!+\!
2\overline{j}_{x,\nu}\! +\! 2\overline{j}_{x-\widehat{\nu},\nu} \big]$ which is an even number. The $I_p(\beta)$
in the weights  ${\cal W}_U$ are the modified Bessel functions and the $P_\phi (2n)$ in 
${\cal W}_\phi$  are the integrals $ P_\phi (2n)  =  \int_0^\infty dr \, r^{2n+1}
\,  e^{-M_\phi^2\, r^2 - \lambda_\phi r^4} = \sqrt{\pi/16 \lambda}  \, (-\partial/\partial M^2)^n \;  
e^{M^4 / 4 \lambda} [1- erf(M^2/2\sqrt{\lambda})]$, which we evaluate numerically and
pre-store for the Monte Carlo. The weight factors $ {\cal
W}_\chi$ are the same as the $ {\cal W}_\phi$, only  the parameters $M_\phi^2$,
$\lambda_\phi$, $\mu_\phi$ are replaced by  $M_\chi^2$, $\lambda_\chi$, $\mu_\chi$. All
weight factors are real and positive. The partition sum (\ref{Zfinal}) thus  is
accessible  to Monte Carlo techniques,  using the plaquette occupation numbers and the
flux variables as the new degrees of freedom.

\vskip2mm
\noindent
{\bf Observables and Monte Carlo update:} In this exploratory study we consider
first and second derivatives of the free energy  as observables
(for   2-point functions and spectroscopy in dual
simulations see,  e.g., \cite{phi4spectro}). In the dual language the observables are mapped 
into weighted sums over dual variables and their fluctuations \cite{DeGaSch1}. 

The dual Monte Carlo update turns out to be rather simple. A detailed
description is given in \cite{DeGaSch1} and here we only introduce the key idea. The algorithm
is a generalization of the worm algorithm \cite{worm} to surfaces with boundaries and we refer
to it as {\it surface worm algorithm} (SWA). The SWA starts with violating the constraints at 
a randomly chosen  link $L_{defect}$ and the two sites at its endpoints by changing the
occupation number of either  the $j$ or the $l$ variable at $L_{defect}$ by $\pm 1$.
Subsequently the occupation number $p$  of one of the six plaquettes attached to $L_{defect}$
is changed such that the violation of the constraint at $L_{defect}$ is healed. Furthermore, for
two of the other links of the plaquette  the constraints are kept intact by changing the $j$
or $l$ fluxes on those links. Thus only at one link of the plaquette the constraints are still
violated and this link is the new defect link $L_{defect}$. Iterating these steps the SWA
propagates the defect link $L_{defect}$ through the lattice until it terminates by inserting a
final unit of $j$ or $l$ flux. Each step of the SWA is accepted with a local Metropolis
decision. We use an additional step for updating
loops of winding $j-l$ flux, and the unconstrained variables $\overline{j}$ and $\overline{l}$ are updated with conventional
Metropolis sweeps.
 
We remark that for checking the correctness of the SWA we compare its results to a local algorithm in the
dual representation \cite{DeGaSch1}
and for $\mu_\phi = \mu_\chi = 0$ also to a simulation in the conventional formulation.

\vskip2mm
\noindent
{\bf Phase diagram at zero density:} We begin with the analysis of the phase diagram and the
properties of the different phases for the case of vanishing chemical potentials $\mu_\phi =
\mu_\chi = 0$. This  also serves as a test of the dual approach which at zero density
can be compared to a conventional simulation. The other parameters are
set to $M_\phi^2 = M_\chi^2 = M^2$ and $\lambda_\phi = \lambda_\chi = 1$ (fixed). 

\begin{figure*}[t]
\begin{center}
\includegraphics[width=59mm,clip]{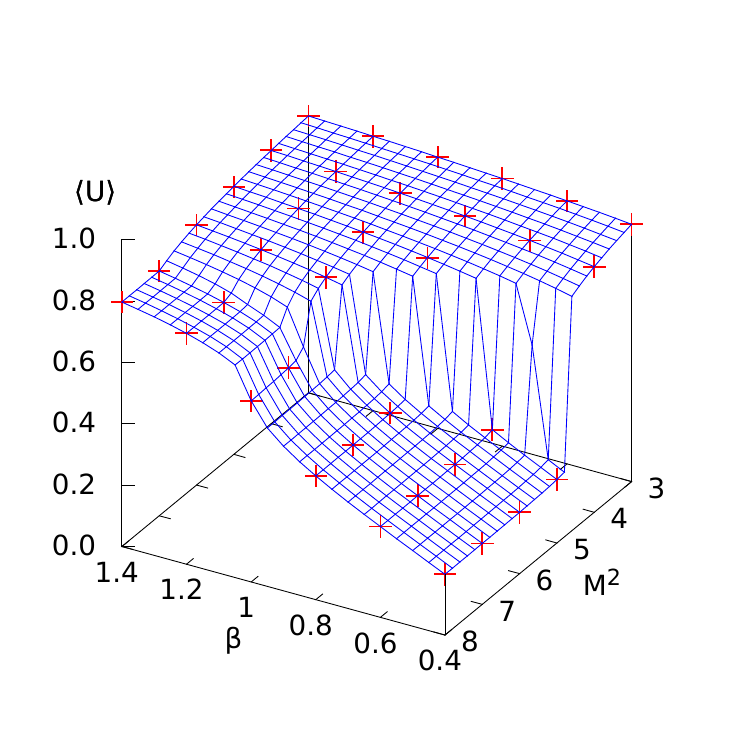}
\hspace{-5mm}
\includegraphics[width=59mm,clip]{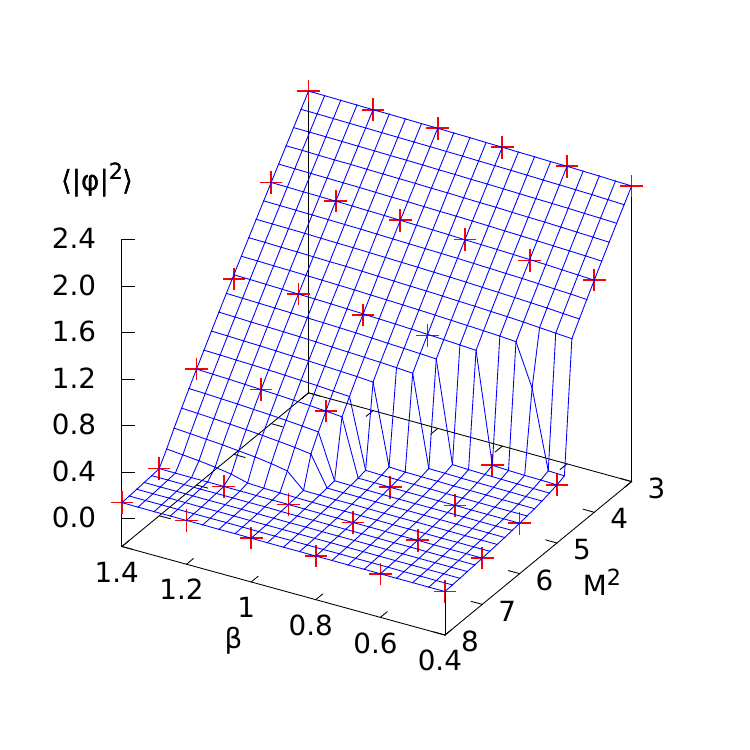}
\hspace{-5mm}
\includegraphics[width=59mm,clip]{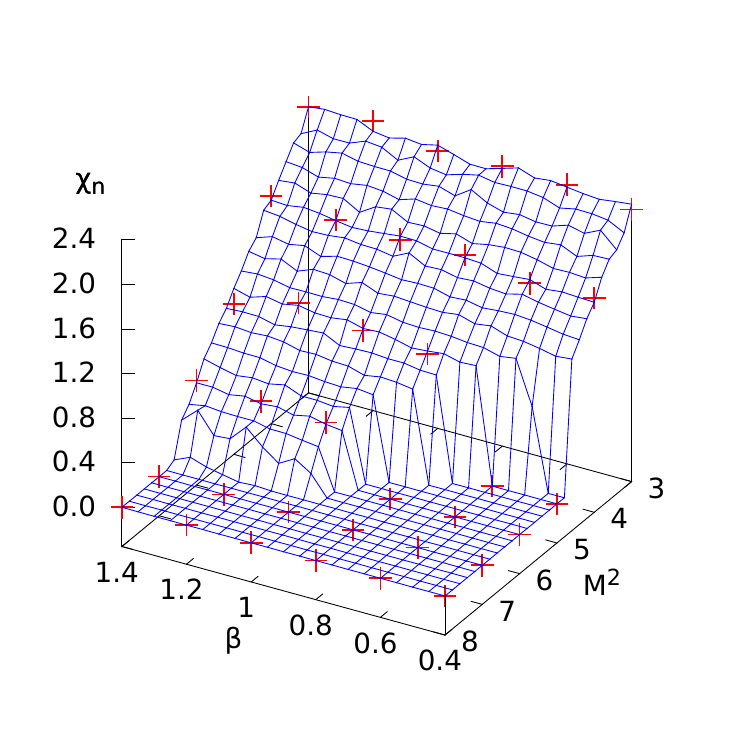}
\end{center}
\vspace{-10mm}
\caption{The observables $\langle U \rangle$, $\langle |\phi|^2 \rangle$ and $\chi_{n_\phi}$ (left to
right) as function
of the inverse gauge coupling $\beta$ and mass parameter $M^2$.}
\label{3Dplots}
\vspace*{-4mm}
\end{figure*}

In Fig.~\ref{3Dplots} we show (left to right) the plaquette expectation value 
$\langle U \rangle$, the expectation value $\langle |\phi|^2 \rangle = 1/V_4 \, \partial \ln Z / \partial
M_\phi^2$ and the particle number 
susceptibility $\chi_{n_\phi} = 1/V_4 \, \partial^2 \ln Z / \partial
\mu_\phi^2$ versus $\beta$ and $M^2$. We remark that at $\mu = 0$
the particle number vanishes, but not $\chi_{n_\phi}$.   

We use results from the dual 
approach for the 3-d mesh in Fig.~\ref{3Dplots} and for some of the parameter values superimpose 
data from a simulation 
in the conventional formulation to check the correctnes of the dual representation (see \cite{DeGaSch1} for
a detailed comparison in the 1-flavor case). 

\begin{figure}[b!]
\centering
\hspace*{-3mm}
\includegraphics[width=75mm,clip]{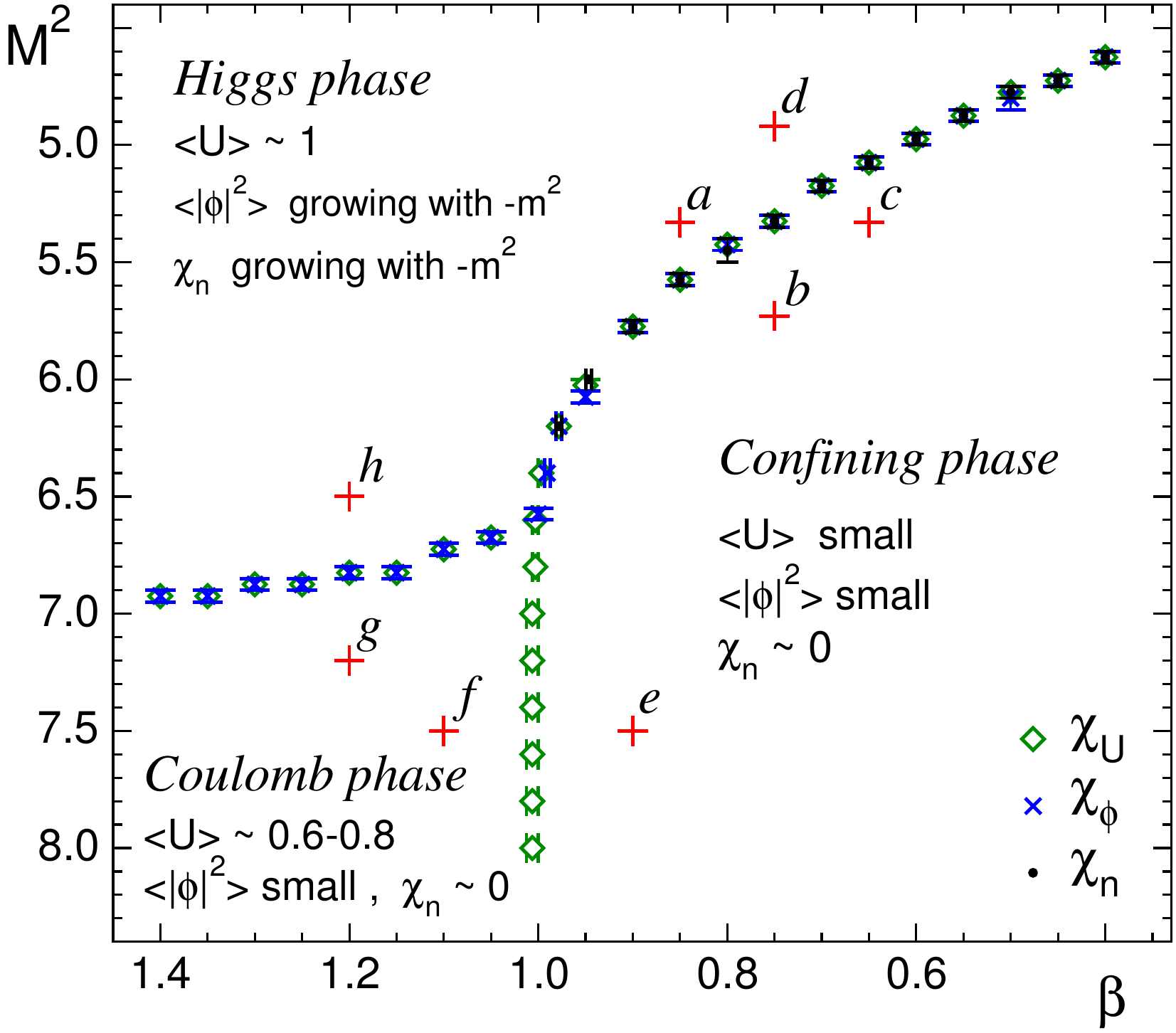}
\caption{Phase diagram in the $\beta$-$M^2$ plane at $\mu = 0$. We show
the phase boundaries determined from the maxima of $\chi_U$ and $\chi_{\phi}$ and the
inflection points of $\chi_n$. We also mark points where we performed runs at finite $\mu$
(plus-signs labelled $a$
to $h$).}
\label{phasediagram}
\end{figure}

For large mass parameter $M^2$ the matter fields decouple and the system is expected to displays the
(very weak) first order transition of pure U(1) lattice gauge theory from the confining to the Coulomb
phase near $\beta \sim 1$. Indeed, for the largest value $M^2 = 8$ we see the behavior of $\langle U
\rangle$ versus $\beta$ as expected  for the pure gauge case, and the critical value of $\beta$ is very
close to 1 (we studied this in more detail using the plaquette susceptibility -- figures not shown).
For vanishing $\beta$ the theory reduces to a charged scalar field, which in the presence of a
$\phi^4$-term has a transition to a Higgs phase.  This strong first order transition is very  
pronounced in all three observables for our smallest coupling  $\beta = 0.4$. It can be located
using the maxima of  the susceptibilities $\chi_U = 1/6V_4 \partial^2 \ln Z / \partial \beta^2$ and  
$\chi_{|\phi|^2} = 1/V_4 \partial^2 \ln Z / ( \partial M_\phi^2)^2$, and  with the
inflection point of $\chi_{n_\phi}$. The result of this analysis is Fig.~\ref{phasediagram}.

The first order line entering our range of parameters at $\beta = 0.4$ and $M^2 \sim 4.6$ (separating Higgs- and confining
phase) shifts towards larger values of $M^2$ for increasing $\beta$ until at $\beta \sim 1.0$, $M^2 \sim
6.6$ the visible  jump in all three observables of Fig.~\ref{3Dplots} vanishes. From that
point on a transition line that separates the confining- and the Coulomb phase continues towards the
first order transition of pure gauge theory discussed above, which is visible in $\langle U \rangle$
and in $\langle |\phi|^2\rangle$ (finer vertical scale is necessary for the latter observable). In addition we
observe a transition line that separates the Higgs- and the Coulomb phase. It connects the
branch point at $\beta \sim 1.0$, $M^2 \sim 6.6$,  to $\beta =
1.4$, $M^2 \sim 6.9$ at the boundary of our parameter range. Thus we can distinguish three phases
characterized by different values of  $\langle U \rangle$, $\langle |\phi|^2 \rangle$
and $\chi_{n_\phi}$ (see the labelling in Fig.~\ref{phasediagram}).

We studied the different transition lines in Fig.~\ref{phasediagram} using finite size analysis of the
second derivatives and histogram techniques, finding that the phase boundary separating  Higgs- and
confining phase is strong first order, the line separating confining- and Coulomb phase is  of weak
first order, and the boundary between Coulomb- and Higgs phase is a continuous transition. 
Our results for the $\mu = 0$ phase diagram are in qualitative
agreement with the conventional results for related
models \cite{Lang}.
 
\vskip2mm
\noindent
{\bf Analysis at finite density:} Let us now come to the case of non-zero $\mu_\phi = \mu_\chi = \mu >
0$. Here the conventional formulation fails and we indeed need the dual approach for obtaining
results. In Fig.~\ref{phasediagram} we mark 8 points (labelled $a$ to $h$) in parameter space where 
we conducted simulations in the range $\mu \in [0,5]$. 

\begin{figure*}[t]
\centering
\hspace*{-5mm}
\includegraphics[width=173mm,clip]{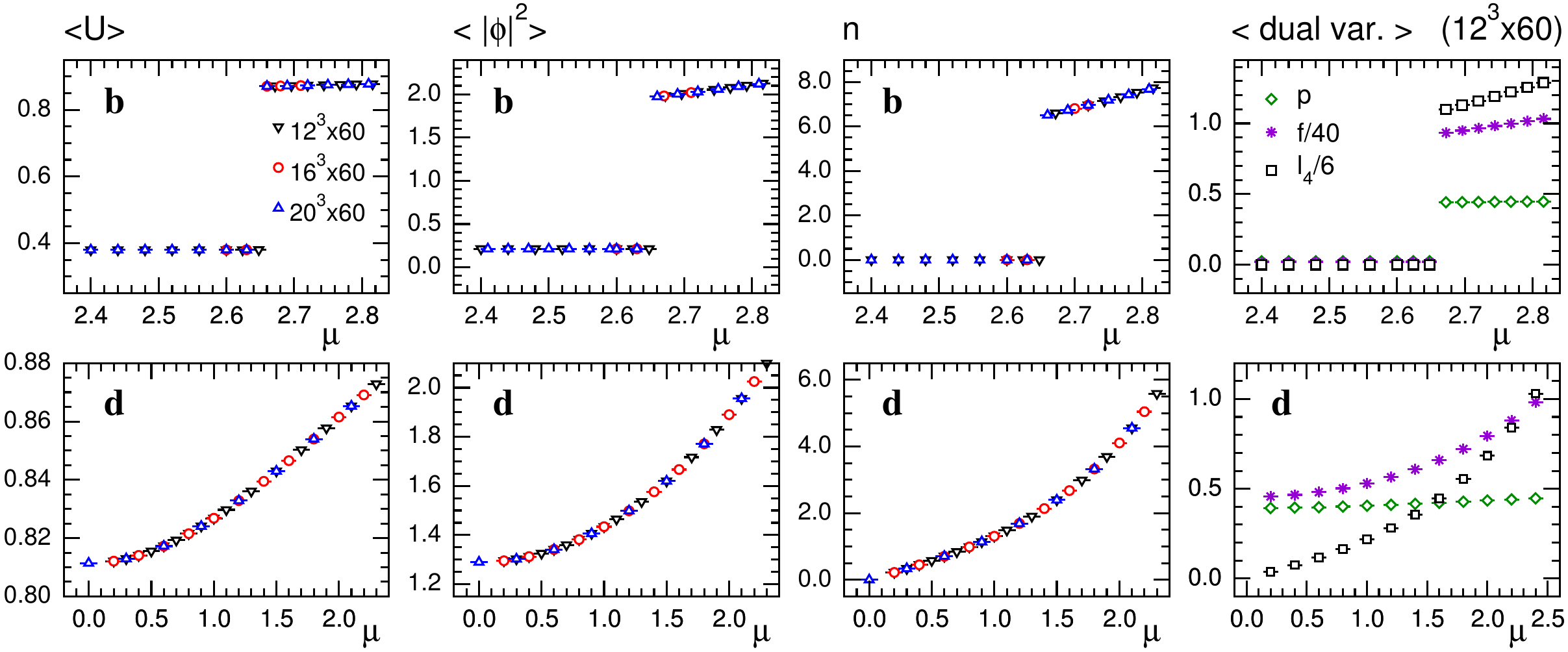}
\caption{From left to right we show the observables $\langle U \rangle$, $\langle |\phi|^2 \rangle$, $n$ and the average dual variables
(normalized with factors as given in the legends) as a function of $\mu$ for points 
$b$ ($\beta = 0.75, M^2 = 5.73$, top row) and $d$ ($\beta = 0.75, M^2 = 4.92$, bottom row).}
\label{pointsbd}
\vspace*{-4mm}
\end{figure*}

 \relax

For 5 of them, points $b$, $c$, $e$, $f$ and 
$g$ we found very similar behavior with a strong first order transition 
as a function of $\mu$ which is visible in $\langle U \rangle$, $n$ and $\langle |\phi|^2 \rangle$. As an 
example in the top row of Fig.~\ref{pointsbd}  we show $\langle U \rangle$, $\langle |\phi|^2 \rangle$ and $n$ as a function of $\mu$ for 
point $b$ ($\beta = 0.75, M = 5.73$).   All three observables jump at $\mu = \mu_c   \sim  2.66$ from values characteristic for the
$\mu = 0$ confining phase to values that correspond to the Higgs phase. We conclude that the finite $\mu$ 
transitions at the points $b$, $c$, $e$, $f$ and $g$ lead into the Higgs phase. This is consistent with the fact, that
finite $\mu$ at tree level changes the mass $m^2 \rightarrow m^2 - \mu^2$, and thus also $M^2 \rightarrow M^2 - \mu^2$. 
This implies that for finite $\mu$ the transition into the Higgs phase takes place for larger values of $M^2$, exactly as we observe.
In other words, the phase boundary of the Higgs phase folds towards larger $M^2$ for increasing $\mu$. 

The points $b$, $c$, $e$, $f$ and 
$g$ have in common that they show a Silver-Blaze type of behavior \cite{cohen} for their finite $\mu$ transition: In the corresponding range of
parameters the $\mu = 0$ theory has a mass gap, and all observables are independent of $\mu$ until $\mu$ reaches the 
value of the mass of the lowest excitation. This behavior is clearly visible in the top row of Fig.~\ref{pointsbd}. Furthermore,
the transition can be seen to be accompanied by a condensation of dual variables. This is obvious from the top plot on the very 
right of Fig.~\ref{pointsbd}, where we show the average plaquette number $p$, the average of $f_x = \sum_\nu\!\big[ |j_{x,\nu}|\!+\!  |j_{x-\widehat{\nu},\nu}|  \!+\!
2\overline{j}_{x,\nu}\! +\! 2\overline{j}_{x-\widehat{\nu},\nu} \big]$ and the average flux $l_4$ in 4-direction 
(normalized with factors as given in the legend). All dual variables jump from very small values to finite numbers at  
$\mu_c   \sim  2.66$.

The situation is different for the points $a$, $d$ and $h$ in the Higgs phase. There we have a Goldstone mode, i.e.,  no mass gap,
and we expect a non-trivial $\mu$-dependence for all values of $\mu$. This behavior is evident in the bottom row plots 
of Fig.~\ref{pointsbd} where 
we show as a prototype example the $\mu$ dependence of the observables when starting from the Higgs phase for point 
$d$ ($\beta = 0.75, M^2 = 4.92$). Here we do not observe discontinuities, but a roughly quadratic behaviour in $\mu$, which can again
be understood from the fact that the observables are essentially linear in $-m^2$ (see Fig.~\ref{3Dplots}) and 
the mass shift $m^2 \rightarrow m^2 - \mu^2$. Also the dual variables show a continuous behaviour and do not condense (bottom row, 
plot at the very right). 

We currently explore the location of the phase boundaries for a wide range of parameters, with the goal of an ab-initio study of 
the various phases suggested for the U(1) gauge-Higgs system at finite $\mu$ \cite{U1GH}.   
 
\vskip2mm
\noindent
{\bf Concluding remarks:} In this exploratory study we have shown that the use of dual variables to overcome the complex action 
problem can be extended also to theories with gauge and matter fields, giving rise to surfaces for the gauge fields 
and loops that bound them for matter. The use of a generalized {\it surface worm algorithm} allows for an efficient update and 
the analysis of the full phase diagram. The same structure of loops and surfaces is expected also for theories where the bosonic matter is replaced by fermions (although with additional minus signs for the loops). This is due to the fact that the fermion determinant can be expanded 
in loops dressed with gauge transporters and integrating out the gauge fields is then done in the same way as here. 
We expect that techniques tested in this paper will be developed further 
and be useful also for other systems with gauge and matter fields.
\vskip2mm
\noindent
{\bf Acknowledgements:} This work is supported by FWF DK 1203 ''Hadrons in Vacuum, Nuclei and Stars'', by DFG SFB TRR55 ''Hadron Physics from Lattice QCD'', by
EU Research Executive Agency (REA), Grant PITN-GA-2009-238353, ITN ''STRONGnet'', and by EU FP7, Grant Nr.~283286, ''Hadron Physics 3''.

\end{document}